\documentstyle[aas2pp4]{article}

\received{3 April 1998}
\revised{}
\accepted{10 September 1998}

\lefthead{U. Klaas et al.}
\righthead{FIR polarisation of the quasar 3C~279}

\begin{document}

\title{Far-infrared polarisation of the quasar 3C~279
      \footnote{Based on observations with ISO, an ESA project with instruments 
         funded by ESA Member States (especially the PI countries: France, 
         Germany, the Netherlands and the United Kingdom) and
         with the participation of ISAS and NASA.}}

\author{Ulrich Klaas\altaffilmark{2}}
\affil{Max--Planck--Institut f\"ur Astronomie, K\"onigstuhl \/17, \/D-69117 
Heidelberg, Germany}
\author{R.J. Laureijs and Jean Clavel}
\affil{ISO Science Operations Centre, Astrophysics Division, Space Science 
Department of ESA, Villafranca, \\ P.O. Box 50727,  E-28080 Madrid, Spain}

\altaffiltext{2}{currently at ISO Science Operations Centre, Astrophysics 
Division, Space Science Department of ESA}

\begin{abstract}
We present the first FIR polarisation results of the OVV quasar 3C~279 obtained 
with
ISOPHOT for two epochs in 1996 and 1997. We describe its integral polarisation 
properties 
at a wavelength of 170 $\mu$m, where the source shows a maximum in its energy 
distribution. 
After a $\gamma$--ray flare in January 1996 a polarisation of  23$\%$ closely 
aligned with 
the radio jet axis was measured in July 1996. In June 1997 the polarisation 
degree had
decreased to 6.5$\%$ with a less good alignment. On the other hand the total 170 
$\mu$m flux 
is the same for both epochs. Our measurements provide additional constraints for 
the 
multi-wavelength properties of synchrotron emission in radio jets and the 
temporal evolution 
of these properties: They show that the FIR radiation of 3C~279 is optically 
thin and that its 
origin is very close to the core. The variability of the FIR polarisation 
without any change 
of the total FIR flux can be explained by a disordering of the magnetic field in 
between the 
core and the first stationary VLBI radio knot.
\end{abstract}


\section{Introduction}

The flat spectrum radio quasar (FSRQ) 3C~279 (z = 0.538; Burbidge and 
Rosenburg\markcite{r2} 1965) is a superluminal radio-source with a one-sided 
radio jet
at a position angle (P.A.) of -140$\arcdeg$ (Unwin et al. \markcite{r26} 1989). 
Its spectral 
energy distribution (SED) extends from the radio to the gamma rays and is
fairly typical of FSRQs. However, observational data are missing in the crucial
mid to far-IR range ($\sim 10^{12} \, to \, 10^{14} Hz$). Simultaneous 
multi-wavelength 
measurements (e.g. Hartman et al.\markcite{r11} 1996) reveal two broad maxima on 
a power per 
logarithmic bandwidth (i.e. ${\rm \nu\,F_{\nu}}$ vs $\nu$) diagram, one in the 
unexplored FIR range and a second one at GeV energies. 

Variations are coherent at all frequencies and the SED ``hardens'' when the 
source 
brightens (Maraschi et al.\markcite{r19} 1994). Flux variations by a factor 
$\sim$~20 and 
$\sim$~5--10 are typical in the GeV and IR-UV-optical ranges, respectively. 
While intra-day variability has been observed in a number of blazars, for 
3C~279 the strongest variability occurs on time scales of a few weeks to $\sim$ 
6 months. 

The optical source is polarised and its polarisation varies both in percentage 
- from 4 to 19\% - and position angle (P.A.) (Angel and Stockman\markcite{r1} 
1980 and 
references therein). During one outburst, simultaneous VLBI and optical 
observations of 3C~279 by Gabuzda et al.\markcite{r7} (1994) show that the 
polarisation in the 
radio regime originates entirely in the jet at a P.A. of 78$\arcdeg$, 
only 10$\arcdeg$ away from the optical polarisation angle (68$\arcdeg$). 
The radio polarisation P.A. is also variable, both in time and spatially 
(Lepp\"anen et al.\markcite{r17} 1995).

FIR observations of 3C~279 out to 200 $\mu$m have only recently become feasible 
with the advent of ISO (Kessler et al.\markcite{r12} 1996) using the photometer 
ISOPHOT 
(Lemke et al.\markcite{r16} 1996). This includes the possibility of performing 
sensitive
far-infrared polarisation measurements of 1--2 Jy sources. Because of the
importance of the infrared wavelength range in the SED of 3C~279 we embarked 
on a program to measure the 170~$\mu$m flux and polarisation of 3C~279 and 
monitor possible variations.

\section{Observations}

Three polarisers constructed of parallel tungsten wires with orientation angles 
of 
$0\arcdeg$, $120\arcdeg$, and $240\arcdeg$ w.r.t. the spacecraft y-axis are 
mounted on
the first change wheel of the ISOPHOT instrument. From tests in the ground 
calibration 
facility (Wolf et al.\markcite{r29} 1994) the actual orientations could be 
verified to be within 
$1\arcdeg$ of the nominal ones.

The observations were done with the C200 camera which consists of 4 individually
stressed Ge:Ga pixels arranged into a 2x2 array. The side length of each pixel 
is 89.4
arcsec with a pixel pitch of 92 arcsec. The 170 $\mu$m filter (technical name: 
C\_160)
was used, being the most sensitive one for the C200 camera.
The bandwidth of an equivalent rectangular bandpass is around 90 $\mu$m with a 
central wavelength
of about 170 $\mu$m (Klaas et al.\markcite{r13} 1994). At 170 $\mu$m the Airy 
disk of a point
source is as large as the whole array, so that for photometric measurements 
point sources are normally centred on the array in order to catch a maximum of 
flux. 
For a higher sensitivity to the peak flux and achievement of a maximum 
source-to-background
contrast it is, however, advantageous to centre the source on one pixel.

As measurement mode the ISOPHOT Astronomical Observation Template PHT51 
(Klaas et al.\markcite{r13} 1994)
was selected. It performs two cycles through the 3 polariser settings of 
$0\arcdeg, 
120\arcdeg, 240\arcdeg$, followed by a photometric measurement and calibrated 
with the internal
Fine Calibration Source (FCS). The time per individual measurement was set to 
128 $s$
during which a good signal stabilisation can be achieved with the C200 camera.
The total performance time per pointing is 1137 $s$ not including the target 
acquisition.
The measurement was repeated on an ``off'' and ``on-target'' position. The 
``off'' position is used for sky background subtraction. 

Table 1 gives the epoch and the positions of the ``off'' and ``on-target''
positions. The ``on-target'' pointing of 1996-07-26 (ISO revolution 253) was 
centred on the C200 array, while on 1997-06-20 (ISO revolution 583) two 
``on-target'' pointings were performed, one again centred on the C200 array
and the other one centred on pixel 4, for the reasons outlined above.

\placetable{tbl-1}

\begin{deluxetable}{lccc}

\tablecaption{Epochs and pointings of 3C~279 observations. \label{tbl-1}}
\tablenum{1}
\tablehead{
\colhead{\bf Position}& \colhead{\bf Date}& \colhead{\bf RA (2000)}& 
\colhead{\bf DEC (2000)}
}
\startdata
  3C~279 off     & 1996-07-26 & $12^{\rm h}56^{\rm m}11\fs1$  & 
              $-05\arcdeg43\arcmin21^{\rm s}$ \nl
  3C~279 on      & 1996-07-26 & $12^{\rm h}56^{\rm m}11\fs1$  & 
              $-05\arcdeg47\arcmin21^{\rm s}$ \nl
  3C~279 off     & 1997-06-20 & $12^{\rm h}56^{\rm m}11\fs1$  & 
              $-05\arcdeg43\arcmin21^{\rm s}$ \nl
  3C~279 on      & 1997-06-20 & $12^{\rm h}56^{\rm m}11\fs1$  & 
              $-05\arcdeg47\arcmin21^{\rm s}$ \nl
  3C~279 pix4    & 1997-06-20 & $12^{\rm h}56^{\rm m}12\fs8$  & 
              $-05\arcdeg48\arcmin21^{\rm s}$ \nl

\enddata

\end{deluxetable}

\section{Data reduction}

ISOPHOT Interactive Analysis (PIA\footnote
{PIA is a joint development by the ESA Astrophysics Division and the ISOPHOT
consortium. The ISOPHOT Consortium is led by the Max--Planck--Institut f\"ur 
Astronomie, 
Heidelberg, contributing ISOPHOT consortium institutes are DIAS, RAL, AIP, MPIK, 
and MPIA.})
V6.5 was used for the standard data reduction up to signal level, including 
ramp linearisation and linear ramp fit, deglitching on ramp and signal level, 
and
dark signal subtraction (Gabriel et al.\markcite{r5} 1996,\markcite{r6} 1997).

For the polarisation analysis we used a specially designed FORTRAN code which
read in the derived signal values and produced all relevant output data based
on the following description (sections 3.2 - 3.5), including the uncertainty 
calculation.

\subsection{Photometry}

The Fine Calibration Source measurement, i.e. the reference measurement against
an instrument internal greybody source calibrated against celestial standards, 
was used to determine the actual detector responsivity (calibration of the photo 
current induced by the incident photons) and the in-band power of the 170 $\mu$m 
filter. 
The resulting in-band power was corrected for the point spread function 
correction 
factors of the C200 array and pixels, respectively (Laureijs et 
al.\markcite{r15} 1998). 
At the time of writing this paper the absolute calibration accuracy of ISOPHOT 
FIR 
photometry is of the order of  30$\%$, the measurement reproducibility is, 
however, 
better than 10$\%$.

\subsection{Determination of Stokes parameters and polarisation values}

For the derivation of the Stokes parameters we used the transformation 
equation for an imperfect polariser as described by formula (20) in Serkowski
\markcite{r22} (1962). In the case of linear polarisation the three Stokes 
parameters $I, Q,
U$ can be derived from the three measured intensities $I_{\rm 0}, I_{\rm 120},
I_{\rm 240}$ in the following way: \\
\begin{equation}
I  =  1/3 * (I_{\rm 0} + I_{\rm 120} + I_{\rm 240})
\end{equation}
\begin{equation}
U  =  1/(\sqrt{3}*P') * (I_{\rm 240} - I_{\rm 120})
\end{equation}
\begin{equation}
Q  =  1/(3*P') * (2*I_{\rm 0} - I_{\rm 120} - I_{\rm 240})
\end{equation}

$P'$ is the polariser efficiency and is assumed to be the same for all three
polarisers. From ground calibration facility measurements $P' = 90\% \pm 3\%$
was determined.

The polarisation degree $P$ and polarisation angle $\theta$ are determined 
from the Stokes parameters in the following way:
\begin{equation}
P  =  1 / I * \sqrt{ U^{\rm 2} + Q^{\rm 2}}
\end{equation}
\begin{equation}
\theta  =  1/2 * arctan(U / Q)
\end{equation}

The polarisation efficiency $P'$ is a scaling factor in $P$ ($P \,\propto\, 
1/P'$), while $\theta$ is independent of it.

\subsection{Sky background subtraction}

Sky background subtraction is performed by subtracting the Stokes parameters 
``off-target'' from the ones ``on-target'' \\
$I_{\rm on} - I_{\rm off} , U_{\rm on} - U_{\rm off} , Q_{\rm on} - Q_{\rm 
off}$. \\
This step is crucial for the finally achieved uncertainty of the polarisation, 
as will be discussed below.

\subsection{Correction for instrumental polarisation}

The instrumental polarisation is derived from measurements of unpolarised 
sources, like stars and background fields, using the same observational set-up.
The average of 14 measurements gives the instrumental polarisation values as 
listed in Table 2. Each C200 pixel has to be seen as a separate entity, as each 
has an individual stressing mechanism, integrating cavity, bias supply and field 
lens. 

\placetable{tbl-2}

\begin{deluxetable}{lcccc}

\tablecaption{Instrumental polarisation determined for the 4 C200 pixels at 
170 $\mu$m. \label{tbl-2}}
\tablenum{2}
\tablehead{
\colhead{\bf C200 pixel} & \colhead{\bf $P_{\rm inst}$} & \colhead{\bf $dP_{\rm 
inst}$\tablenotemark{b}} & 
\colhead{\bf $\theta_{\rm inst}$\tablenotemark{a}} & \colhead{\bf $d\theta_{\rm 
inst}$\tablenotemark{b}} \\
\colhead{} & \colhead{[$\%$]} & \colhead{[$\%$]} & \colhead{[$\arcdeg$]} & 
\colhead{[$\arcdeg$]}
}
\startdata
  1  &  4.9  &  0.3  &  -57.1  &  0.8  \nl
  2  &  5.5  &  0.3  &  +87.7  &  1.1  \nl
  3  &  6.1  &  0.3  &  -39.5  &  0.8  \nl
  4  &  3.4  &  0.3  &  -76.5  &  1.8  \nl

\enddata

\tablenotetext{a}{The angles are counted w.r.t. the ISO spacecraft (S/C) y-axis}
\tablenotetext{b}{The quoted uncertainties are the standard deviations of the 
mean}

\end{deluxetable}

The instrumental polarisation is corrected for on the level of the normalised 
Stokes parameters $u = U/I , q = Q/I$ in the following way:
\begin{equation}
u' = u - P_{\rm inst} * sin(2 * \theta_{\rm inst})
\end{equation}
\begin{equation}
q' = q - P_{\rm inst} * cos(2 * \theta_{\rm inst})
\end{equation}

\subsection{Calculation of uncertainties}

The calculation of uncertainties was done on all levels of the data reduction 
described above.
Both Gaussian and maximum error propagation were performed in order to judge the
robustness of the final results. For each formula the partial derivatives were
calculated and summed up quadratically (Gaussian uncertainty) or absolutely 
(maximum uncertainty).
In this paper we only quote the Gaussian uncertainties. The derived maximum 
uncertainties are by
factors 2.5 - 4 higher than the Gaussian ones.

For the uncertainty of the signal determination the reproducibility of the 
signal in
the different polariser cycles was used as a representative measure. A kind of 
standard
deviation was constructed by deriving the average signals for the three 
polariser 
settings $\bar{I_{\rm 1}}$, $\bar{I_{\rm 2}}$, $\bar{I_{\rm 3}}$ and determining

\begin{equation}
   dI = 1 / \sqrt{3 * n_{\rm cyc} - 1} *
        \sqrt {\sum\limits_{i=1}^3 \sum\limits_{j=1}^{n_{\rm cyc}} 
        (I_{\rm i,j} - \bar{I_{\rm i}})^{\rm 2}}
\end{equation}

For all our measurements $n_{\rm cyc}$ was equal to 2.
For sky measurements a similar uncertainty from the sky background measurement 
has to be
taken into account. For instrumental polarisation measurements no background 
subtraction
was done, but the total sky flux was considered to be unpolarised.

Other uncertainty factors taken into account are the 3$\%$ uncertainty of the 
polarisation
efficiency and two times the standard deviation of the mean instrumental 
polarisation
as quoted in Table 2.

As a residual uncertainty in the Stokes parameters $Q$ and $U$ always yields an 
effective
positive polarisation degree, this can be corrected for in the case of low S/N 
ratio 
following a method as described by Wardle and Kronberg \markcite{r27} (1974) and 
by 
determining the so-called de-biased polarisation degree:
\begin{equation}
P_{\rm debias} = \sqrt{P^{\rm 2} - dP^{\rm 2}}
\end{equation}

This uncertainty calculation showed clearly the reliability of the ISOPHOT FIR 
polarisation
measurements, but also the limitations of the mode. Signal transients impacting 
on the
reproducibility proved to be on a very low level of 0.3$\%$ - 2$\%$. The most 
limiting
parameter is the source--to--background contrast. With a ratio of 2:1 
source/background
5$\%$ polarisation can be reliably measured on a 3 - 5 $\sigma$ level. With a 
ratio of
0.3:1 source/background the uncertainty becomes as high as 5$\%$.

\section{Results}

Table 3 lists the polarisation results per pixel. It can be seen that the second 
measurement
on 1997-06-20 has a significantly smaller uncertainty due to centring the source 
on the 
detector pixel. The detection of polarisation in the measurement centred on the 
array is 
only marginal. The source--to--background contrast is about 0.37 for the 
measurement centred 
on the array versus 1.54 for the measurement centred on the pixel. The 
measurement on 
1996-07-26 had even lower source-to--background contrast, on average only 0.23, 
as it had been 
performed at smaller solar elongation yielding a stronger Zodiacal Light 
contribution. 
Despite this fact, the consistently high polarisation degree found for all 4 
pixels gives 
a similar final uncertainty (of the mean), cf. Table 4.

\placetable{tbl-3}

\begin{deluxetable}{lccccccc}

\tablecaption{Polarisation results for 3C~279 from the two epochs in 1996/97 
(per C200 pixel). \label{tbl-3}}
\tablenum{3}
\tablehead{
\colhead{\bf date} & \colhead{\bf pix} & \colhead{\bf S/N\tablenotemark{a}} & 
\colhead{\bf $P$} & 
\colhead{\bf $dP$\tablenotemark{b}} & \colhead{\bf $\theta$\tablenotemark{c}}  & 
\colhead{\bf $d\theta$\tablenotemark{b}} & \colhead{\bf $\phi_{\rm 
S/C}$\tablenotemark{d}} \\
\colhead{} & \colhead{} & \colhead{} & \colhead{[$\%$]} & \colhead{[$\%$]} & 
\colhead{[$\arcdeg$]} & 
\colhead{[$\arcdeg$]} & \colhead{[$\arcdeg$]}
}

\startdata
1996-07-26 &   1   &   25   &   27.2   &   4.1     &   60.2      &     4.9      
&    22.6    \nl
1996-07-26 &   2   &   17   &   21.1   &   6.0     &   64.8      &     8.4      
&    22.6    \nl
1996-07-26 &   3   &   30   &   24.0   &   3.2     &   50.4      &     5.0      
&    22.6    \nl
1996-07-26 &   4   &   30   &   20.3   &   3.2     &   49.4      &     6.3      
&    22.6    \nl
1997-06-20 &   1   &   11   &   10.1   &   8.3     &   81.4      &    22.8      
&    22.8    \nl
1997-06-20 &   2   &   11   &   10.4   &   8.8     &   76.2      &    22.7      
&    22.8    \nl
1997-06-20 &   3   &   22   &    4.5   &   5.0     &   79.5      &    28.1      
&    22.8    \nl
1997-06-20 &   4   &   19   &    3.2   &   4.8     &   41.1      &    56.3      
&    22.8    \nl
1997-06-20 &   4   &   83   &    6.6   &   1.5     &   75.2      &     5.6      
&    22.8    \nl

\enddata

\tablenotetext{a}{Uncertainties are 1 $\sigma$ values}
\tablenotetext{b}{For the source only, including the uncertainty by background 
subtraction}
\tablenotetext{c}{Angle w.r.t. the S/C coordinate frame (y-axis)}
\tablenotetext{d}{Angle of S/C coordinate frame w.r.t. the sky N-direction}

\end{deluxetable}

The final polarisation degree and angle as listed
in Table 4 was derived by applying the debiasing correction as described in the 
section
``calculation of uncertainties'' and averaging the values of the 4 pixels for 
the first measurement
(weighted mean), as well as adding the angle of the S/C y-axis on the sky. For 
the marginal
polarisation detection of the measurement centred on the array in 1997-06-20 we 
give only
the weighted mean without any debiasing, as two pixels would give a ``zero'' 
result.

Table 4 also contains the flux and its uncertainty as derived for each epoch. 
The two 
measurements centred on the array are compared best, because the beam conditions 
are
the same. They give a remarkable equality of the total flux for the two epochs, 
but
also the measurement centred on pixel 4 is within the uncertainty consistent 
with an equal
flux at the two epochs. Photometric measurements by Haas et al.\markcite{r10} 
(1998) determined a
170 $\mu$m flux of 1.5 $Jy$ on 1996-12-19 with a thermal emission bump showing 
up on top of
the synchrotron emission. The equality of flux at our two measurement epochs 
seems therefore
to be accidental: In the period July 1996 -- December 1996 the non-thermal flux 
was decreasing
towards a minimum, while in June 1997 it was rising again.

\placetable{tbl-4}

\begin{deluxetable}{lcccccc}

\tablecaption{Integral FIR flux and polarisation properties of 3C~279 for 
1996-07-26 and 1997-06-20. 
\label{tbl-4}}
\tablenum{4}
\tablehead{
\colhead{\bf Julian} & \colhead{\bf $F_{\rm \nu}$} & \colhead{\bf $dF_{\rm 
\nu}$} & \colhead{\bf $P$}  & 
\colhead{\bf $dP$} & \colhead{\bf $\theta$\tablenotemark{a}}  & \colhead{\bf 
$d\theta$} \\
\colhead{\bf date} & \colhead{[Jy]} & \colhead{[Jy]} & \colhead{[$\%$]} & 
\colhead{[$\%$]} & 
\colhead{[$\arcdeg$]} & \colhead{[$\arcdeg$]}
}

\startdata
2450290.703  &   2.51     &      0.22      &  22.8      &   1.6      &   77.9    
      &     3.0         \nl
2450619.509  &   2.52     &      0.11      &  (5.4)     &   \nodata  &   (99.5)  
      &     \nodata     \nl
2450619.522  &   2.43     &      0.14      &   6.5      &   1.5      &   98.0    
      &     5.6         \nl

\enddata

\tablenotetext{a}{Polarisation angles are given w.r.t. the sky N-direction}

\end{deluxetable}

\section{Discussion}

All the observational characteristics outlined in the introduction are 
consistent 
with a model where a relativistic jet pointing very close to the line-of-sight 
dominates the emission of 3C~279: the IR-optical-UV 
flux represents the optically thin synchrotron emission of the relativistic 
electrons, while X-rays and gamma-rays are produced by inverse compton (IC) 
up-scattering of ambient ``seed'' photons off the same population of electrons 
(e.g. Sikora\markcite{r23} 1994). Low frequency synchrotron self-absorption in a 
compact 
inhomogeneous source such as 3C~279 naturally gives rise to the observed flat 
radio spectrum. It is not yet known whether the ``seed'' photons which are 
up-scattered to high energies are the synchrotron photons themselves (the 
so-called synchrotron self-Compton or SSC model) or external thermal photons 
from e.g. the accretion disk (the external Compton or EC model). In 
3C~279, however, the relative amplitude of the variations of the IC and 
synchrotron components slightly favors the SSC model over the EC model 
(Maraschi et al.\markcite{r19} 1994). A high degree of linear polarisation is a 
natural 
signature of the synchrotron process.

A significant thermal contribution to the 170 $\mu$m flux, in particular by dust 
in the host galaxy, can be ruled out. Even if the quasar host belonged to the 
class of ultra-luminous IR galaxies (ULIRGs), its expected contribution 
(Klaas et al.\markcite{r14} 1997) at the distance of 3C~279 would be $\leq 0.3 
Jy$, i.e. 
about 10$\%$ of the measured 170 $\mu$m flux, and can be safely neglected. This 
is
consistent with the SED composition as described by  Haas et al.\markcite{r10} 
(1998) 
who found that the synchrotron component is the dominant one at 170 $\mu$m.

There was no further flaring episode of 3C~279 after the $\gamma$--ray flare of 
January 1996 reported in IAU circular 6294 (Wehrle and Hartman\markcite{r28} 
1996). 
Hence, it is meaningful to compare our FIR flux to the average 0.8 mm flux over 
the period
1988--1994, $\simeq~10~Jy$ (Stevens et al.\markcite{r24} 1994). In a power per 
logarithmic bandwidth (i.e. ${\rm log\,\nu\,F_{\nu}}$) diagram, with the flux 
expressed in $Jy$ and the frequency in $Hz$, this corresponds to 12.57. Our 
measurements at 170~$\mu$m yield 12.65 both in July 96 and June 97. 
This value is higher than the mm flux and lies along the 
extrapolation of the radio-mm spectrum (fig. 1 of Maraschi et al.\markcite{r19} 
1994). This is a 
strong indication that the synchrotron ``break'' occurs at wavelengths shorter 
than 170~$\mu$m, i.e. ${\rm \nu_{s}\,\geq\,1.76\,10^{12}}~Hz$, at least when 
3C~279 is not in outburst. This lower limit agrees with the
intersection of the extrapolated optical-IR synchrotron power-law and the
extrapolated cm-mm power-law of 3C~279 in a quiescent state, 
${\rm \nu_{s}\,\simeq\,3.5\,10^{12}}~Hz$ (Litchfield et al.\markcite{r18} 1995). 

Determining precisely the ``break'' frequency $\nu_s$ of the peak of the 
synchrotron component is of particular importance: in the SSC model, when 
combined with the peak-frequency of the IC component $\nu_c$, it yields 
directly the maximum energy exponent $\gamma_b$ of the relativistic electrons in 
the 
jet: ${\rm \gamma_b\,\simeq\,(\frac{3\,\nu_c}{4\,\nu_s})^{1/2}}$. Using
${\rm \nu_c\,\simeq\,10^{23}}~Hz$ (Ghisellini et al.\markcite{r8} 1996), we 
obtain 
${\rm \gamma_b\,\leq\,2.1\,10^5}$. The knowledge of both the frequencies and the 
luminosities 
of the synchrotron and self-Compton peaks and the variability time-scale 
constrains the 
parameters of the SSC model, in particular the magnetic field and the Doppler 
factor 
(Ghisellini et al.\markcite{r8} 1996).

The FIR polarisation angle we measure in July 1996 agrees to within 2$\arcdeg$
with the 1.1 and 0.8 mm polarisation angle measured by Stevens et 
al.\markcite{r25} (1996) in 
August 1995 ($76\pm2\arcdeg$) and is also close to the optical polarisation 
angle 
(68$\arcdeg$) as measured by Gabuzda and Sitko\markcite{r7} in 1987 while 3C~279 
was in 
outburst. The inner base of the 6 cm VLBI mas jet lies at  P.A.~=~-120$\arcdeg$ 
(Cawthorne and Gabuzda\markcite{r4} 1996) and is therefore also roughly aligned 
with the 
FIR and mm polarisation angle. To a first approximation, the magnetic field 
direction as determined from mm and FIR measurements is thus perpendicular 
to the base of the jet. The same conclusion was reached by Nartallo et 
al.\markcite{r20} 
(1998) from mm polarisation measurements at eleven epochs between 1991.43 and 
1995.98. 
These authors find that the 1.1 mm P.A. varies between 31$\arcdeg$ and 
76$\arcdeg$ around a 
mean value of 50$\fdg$5 with an r.m.s. dispersion of 12$\fdg$5. 
This contrasts with the radio polarisation
data (epoch 1987 May) which show that the magnetic field is roughly aligned with 
the
VLBI jet at cm wavelengths (Cawthorne and Gabuzda\markcite{r4} 1996). As 
suggested by 
Stevens et al.\markcite{r25} (1996), this difference can be understood as 
follows: 
on the one hand, the synchrotron emission is optically thin at IR and mm 
frequencies and therefore dominated by the most energetic electrons 
at the base of the jet, while radio emission includes a larger relative 
contribution from the thinner knots at some distance along the jet than from 
the largely saturated optically thick radio core. On the other hand, it has 
been established that, for quasars, the {\em longitudinal\/} component of 
the magnetic field increases in strength with distance from the core, 
presumably because the field is aligned by shear (Cawthorne et al.\markcite{r3} 
1993). 
On these grounds, one expects a 90$\arcdeg$ change in polarisation P.A. from 
radio to FIR wavelengths, as indeed observed. The large drop in polarisation 
in eleven months we measure is also consistent with a compact origin 
(i.e. the core) for the bulk of the FIR polarised flux.

The maximum degree of linear polarisation of an optically thick synchrotron
source is 12$\%$, while the maximum polarisation of an optically thin 
synchrotron 
source is about 80$\%$ (Pacholczyk\markcite{r21} 1970). Hence, the high degree 
of 
polarisation we measure in July 1996 confirms that the 170~$\mu$m emission was 
optically thin. In June 1997, the degree of polarisation is formally compatible 
with
the optically thick case but only for unrealistically low values of 
${\rm \gamma_b\,\leq\,6}$. The very high upper limit 
on ${\rm \gamma_b}$ derived previously suggests that the FIR emission was
optically thin in June 1997 as well.

The factor 3.5 drop in polarisation and 20$\arcdeg$ shift in P.A. between July 
1996 and June 1997 is difficult to reconcile with the equality of the 
FIR flux. Our result is different from that of Nartallo et al.\markcite{r20} 
(1998) 
whose data showed a positive correlation between the polarisation percentage 
at 1.1 mm and the 1.1 mm flux. Such a correlation is expected in a shock model 
where 
compression of the magnetic field leads to an increase of both flux and 
polarisation. 
The behaviour in the FIR does not fit that pattern. As noted above, the 
170~$\mu$m flux 
we measure is consistent with 
an extrapolation of the radio spectrum when the jet is in a {\em nearly 
quiescent\/} state
(definitely not in outburst). Hence, 
it is perhaps not surprising that the shock model does not apply in such 
circumstances. The 
decrease in FIR polarisation at equal flux level could possibly be explained as 
the 
result of the varying contributions from different jet components.

The 22~GHz VLBA map of 3C~279 by Lepp\"anen et al.\markcite{r17} (1995) shows a 
complex polarisation structure: close to the core, the fractional polarisation 
is $\sim~13\%$ and the P.A. is roughly aligned along the jet axis; 1 mas 
away, at the location of the C5 stationary knot, the fractional polarisation 
drops to a minimum of $10 \pm 4~\%$ and flips by 90$\arcdeg$. Further away, the 
polarisation increases again to a maximum of $33 \pm9~\%$ and then decreases 
again at the location of the C4 knot (2 mas) to a value of $15 \pm 1~\%$ while 
re-orienting itself parallel to the jet. Hence, the drop in polarisation and 
shift in P.A. between 1996 and 1997 could in principle be explained by 
geometrical arguments: in July 1996, the FIR flux was dominated by a recently 
formed knot close to the core (possibly the aftermath of the January 1996 
gamma-ray flare); assuming it propagates at near light speed, in June 1997 the 
shock could have travelled at most $\sim~4~\%$ of the distance (since the jet is
seen in projection) to the C5 knot 
(8 pc; ${\rm H_{0}\,=\,75~km\,s^{-1}\,Mpc^{-1}}$; ${\rm q_{0}\,=\,0.5}$). 
If, at such a distance, the magnetic field retains the same intensity but
is already in the process of re-orienting itself and acquiring the geometrical
configuration it has at the location of the C5 knot, it could become disordered.
A disordered field would explain the observed drop in polarisation and 
shift in P.A. with no change in the synchrotron emissivity. This explanation 
is a bit contrived however; more observations are definitely required to 
better constrain the jet model. Unfortunately, after 2nd August 1997, 3C~279
was no longer visible for ISO due to pointing constraints w.r.t. the Sun and
the Earth limb.

\section{Conclusion}

We measured for the first time the far-infared polarisation of a quasar. With
ISOPHOT 170 $\mu$m polarisation degrees of $\geq 5\%$ can be reliably measured, 
with
an accuracy of 1.5$\%$, for sources as faint as 2 $Jy$.

For 3C~279 in particular we found a strong variability in the 170 $\mu$m 
polarisation 
degree and to some extent also a variation in the polarisation angle. In July 
1996, 
about 6 months after a $\gamma$--ray flare, a high polarisation degree of about 
23$\%$ and 
aligned within better than 20$\arcdeg$ with the direction of the VLBI radio jet 
was measured. 
11 months later the polarisation degree had decreased to 6.5$\%$ and the 
polarisation angle had
rotated away from the VLBI radio jet direction by about 40$\arcdeg$. At the two 
epochs
the total flux of the source was equal, but from independent photometric 
measurements it is
known that the total flux at 170 $\mu$m went through a minimum of about 1.5 $Jy$ 
in December 1996.

Our polarisation results imply that the FIR synchroton radiation is optically 
thin and
emerges from very close to the core. The observed variation in polarisation
with equality of the total flux could be explained by a disordering of the
magnetic field inbetween the core region and the first stationary radio knot.

\acknowledgments
The ISOPHOT FIR polarisers were provided by the co--investigators E. Kreysa and 
H.--P. Gem\"und from Max--Planck--Institut f\"ur Radioastronomie, Bonn, Germany. 
ISOPHOT was funded by the Deutsche Agentur f\"ur Raumfahrtangelegenheiten DARA, 
the Max--Planck--Society, the Danish, British and Spanish Space Agencies and 
several European and American Institutes. Our thanks go also to all ESA ISO 
Operational 
Teams and ISOPHOT consortium institutes supporting the commissioning of the 
ISOPHOT
observation modes, in particular to the colleagues of the ISOPHOT Data Centre at
MPIA Heidelberg who provided us also with the results of the polarisation tests
performed in the ground calibration facility. We thank Paul Barr for initiating 
the
project of 3C~279 polarisation measurements. Special thanks to Michael 
Linden--V{\o}rnle 
from DSRI for help in the early polarisation commissioning and Ingolf 
Heinrichsen from PIDT
for help in reduction of some of the data presented here. For literature search 
we used the NASA/IPAC Extragalactic Data Base (NED). We thank the referee for 
constructive
comments and pointing out to us the publication of the 1.1 mm polarisation 
monitioring measurements.

\end{document}